\documentstyle[aps,manuscript]{revtex}

\newcommand{\be}{\begin{equation}}
\newcommand{\ee}{\end{equation}}
\def\la{\mathrel{\mathpalette\fun <}}

\def\fun#1#2{\lower3.6pt\vbox{\baselineskip0pt\lineskip.9pt
\ialign{$\mathsurround=0pt#1\hfil##\hfil$\crcr#2\crcr\sim\crcr}}}
\begin{document}
\def\baselinestretch{1.5}
\normalsize

\title{Meson Masses in Nuclear Matter}
\author{V. L. Eletsky$^{1,2}$ and B. L. Ioffe$^{1,3}$}
\address{$^1$Institute of Theoretical
and Experimental Physics,
 B.Cheremushkinskaya 25, Moscow 117259, Russia}
 \address{$^2$Institut f\"ur Theoretische Physik III, Universit\"at
Erlangen-N\"urnberg, D-91058, Erlangen, Germany}
 \address{$^3$Institut f\"ur Kernphysik, Forschungszentrum J\"ulich, D-52425,
 J\"ulich, Germany}

\maketitle
\begin{abstract}
Mass shifts $\Delta m$ of particles in nuclear matter relative
to their vacuum values are considered. A general formula relating
$\Delta m(E)$ ($E$ is the particle energy) to the real part of the forward
particle-nucleon scattering amplitude ${\rm Re} f(E)$ is presented and
its applicability domain is formulated. The $\rho$-meson mass
shift in nuclear matter is calculated at $2\la E_{\rho}\la 7$ GeV
for transversally and longitudinally polarized $\rho$-mesons
with the results:  $\Delta m_{\rho}^T \sim 50$ MeV and $\Delta m_{\rho}^L
\sim 10$ MeV at normal nuclear density.
\end{abstract}

\vspace{2mm}
\vspace{2mm}
The problem of how the properties of mesons and baryons change in nuclear
matter in comparison to their free values has attracted a lot of attention
recently. Among these properties the first of interest are mass
shifts of particles in nuclear matter. This interest is related to the fact
that it was possible to calculate by QCD sum rules
and on the lattice almost all masses of low lying mesons and baryons, and a
hope appears to extend these calculations to the case of particles is
embedded in nuclear medium. On the other hand, the values of particle masses
can be measured experimentally - at least some of them - and some data
started to appear. In this aspect experiments on heavy
ion collisions, in which the dependence of particle masses on nuclear
density can be found, are very promising.

In early theoretical investigations of this problem\cite{1,2} one or
another model of strong interaction of particles in nuclear matter was used.
In the pioneering work by Drukarev and Levin\cite{3} the use of QCD sum
rules for the calculation of nucleon mass shift in nuclear matter was
suggested. Later this method was applied also to calculation of meson
masses (for recent reviews see\cite{4,5,6}). Among the latter the most
interesting is the case of light vector mesons and, especially, of $\rho$.
The reason is that a $\rho$-meson produced inside the nucleus decays also
there and can be observed by its partial decay into $e^{+}e^{-}$. So, the
characteristics of $\rho$-meson inside the nucleus can be directly measured.

The masses of vector mesons in nuclear matter were calculated
in\cite{2,7,8,9,10,11,12,13}. (In Ref.\cite{9} a universal ratio of particle
masses in nuclear matter to their vacuum values was suggested.) However,
the results obtained by different methods do not coincide. Moreover, there
is no agreement as to whether the $\rho$ mass decreases or increases in
nuclear medium in comparison with its value in vacuum: in
Refs.\cite{2,7,8,12} $\Delta m = (m_{\rho})_{nucl} - m_{\rho}> 0$,
while in Refs.\cite{9,10,13} $\Delta m < 0$. Since the interaction of
$\rho$-meson with nucleons in medium is energy dependent, one may expect
that the mass shift is also energy dependent. This problem was not
considered in the investigations mentioned above: only the case of
$\rho$-meson at rest was considered. But $\rho$-mesons at rest are not good
objects from experimental points of view. In experiments on nuclei
$\rho$-mesons as a rule are produced with energies of order of $1$ GeV or
more. Finally, for a moving $\rho$-meson the interaction with matter of
the meson polarized transversally or longitudinally
is different. So, one may expect that the mass shifts in nuclear matter of
transversally and longitudinally polarized $\rho$-mesons are also different.
For all of these reasons a new consideration of this problem is desirable.

We start with general considerations applicable to any particle imbedded in
nuclear matter. Let us accept the standard assumption in the treatment of
the problem in view\cite{4,5,6}: the interaction of the particle with a
nucleon in matter is not affected by other nucleons, i.e. the nuclear matter
can be considered as an inhomogeneous macroscopic medium. This immediately
restricts the particle wave length: $\lambda =k^{-1}\ll d$, where $d$ is the
mean internucleon distance. Numerically this means
that the particle momentum $k$
must be larger than a few hundred MeV. Since we assume
that the particle is created
inside the nucleus, we must require that its formation length $l_{form}$ is
less than the nucleus radius $R$

\be
l_{form}\sim\frac{E}{m}\frac{1}{m_{char}}\; ,
\label{lform}
\ee
where $E$ and $m$ are the particle energy and mass,
$m_{char}\sim m_{\rho}$ is the
characteristic strong interaction scale.
Eq.~(\ref{lform}) implies an upper limit on the particle energy,
$E_{\rho}< 15$ GeV for middle weight nuclei. An additional restriction
on the upper value of the particle momentum $k$ arises from the requirement
that for the observation of the mass shift the particle must mainly decay
inside the nucleous, $k/\Gamma m< R$.
This gives $k_{\rho}< 6$ GeV, $k_{\omega}< 300$ MeV,
$k_{\phi}< 200$ MeV for $\rho$, $\omega$, and $\phi$, correspondingly.
Comparison of lower and upper limits for the particle momenta shows that for
$\omega$ and $\phi$ the assumption of independent scattering on individual
nucleons in the nucleous and the possibility of experimental observation are
in contradiction. So, we are left only with $\rho$.

In consideration of the particle mass shifts in nuclear matter, or,
equivalently, of the mean effective potential acting on the particle in
matter, we use the general method suggested long ago for treatment of
propagation of fast neutrons in nuclei\cite{14} (see also\cite{15}). The
main idea is that for $\lambda\ll d\ll R$ the effect of medium on the
particle propagation can be described by attenuation
and refraction indeces. Attenuation of particles moving in the direction of
$z$-axis at a distance $z$ is equal to
$\exp (-\rho\sigma z)$, where $\rho =A/V$
is the nuclear density, $A$ is the atomic number, $V$ is the nucleus
volume, and $\sigma$ is the total cross section of the interaction of the
particle with nucleons. (Strictly speaking $\rho\sigma =(Z\sigma_p +
N\sigma_n)/V$.) Using the optical theorem

\be
k\sigma = 4\pi {\rm Im} f(E)\; ,
\label{opt}
\ee
where $f(E)$ is the forward scattering amplitude, we can write that the
modulus of the particle wave function in matter is proportional to

\be
|\psi|\sim\exp\left[ -\rho\frac{2\pi z}{k} {\rm Im} f(E)\right]
\label{psi}
\ee
This formula is evidently generalized to the wave function itself

\be
\psi\sim\exp\left[i\rho\frac{2\pi z}{k} f(E)\right]
\label{psi1}
\ee
Eq.~(\ref{psi1}) is correct if $|f|\ll d=(V/A)^{1/3}$: only in this case the
scattering on each nucleon can be considered as independent and interference
effects can be neglected\cite{15}. ${\rm Re} f(E)$ is related to the
refraction index of matter for particle propagation\cite{14}.  We want to
decribe the propagation of a particle through nuclear matter introducing an
effective mass $m_{eff}=m+\Delta m$.  This means that (leaving absorption
aside)

\be
\psi\sim e^{ikz}\; ,~~~k=\sqrt{E^2 -m_{eff}^2}\approx k-\frac{m}{k}\Delta m
\label{psi2}
\ee
By comparing Eqs.~(\ref{psi1}) and (\ref{psi2}) we get

\be
\Delta m (E) = -2\pi\frac{\rho}{m}{\rm Re} f(E)
\label{dm}
\ee
The expression in Eq.~(\ref{dm}) for $\Delta m$ has the meaning of an
effective potential acting on the particle in medium\cite{14,15}.
For the correction to the particle width we have in a similar way

\be
\Delta\Gamma (E) =\frac{\rho}{m}k\sigma (E)
\label{w}
\ee
All the above statements are general and can be applied to any particle in
nuclear matter. Let us now turn to the most interesting case, the
$\rho$-meson.

In order to find $\rho N$ forward scattering amplitude we use the vector
dominance model (VDM) and the relation which follows from VDM
(see, e.g.\cite{16})

\be
f_{\gamma N}=4\pi\alpha\left(
\frac{1}{g_{\rho}^2}f_{\rho N}+\frac{1}{g_{\omega}^2}f_{\omega N}+
\frac{1}{g_{\phi}^2}f_{\phi N}\right)
\label{vdm}
\ee
The last term in the r.h.s. of Eq.~(\ref{vdm}) can be safely neglected: as
follows from $\phi$-photoproduction data, it is small. Basing on the quark
model, assume $f_{\omega N}\approx f_{\rho N}$. Since
$g_{\omega}^2/g_{\rho}^2\approx 8$, the contribution of $\omega$ to the
r.h.s. of Eq.~(\ref{vdm}) is also small. Therefore, according to
Eq.~(\ref{vdm}) ${\rm Re} f_{\rho N}(E)$ is expressed through
${\rm Re} f_{\gamma N}(E)$. The latter can be found from the
photoproduction data through the dispersion relation with one subtraction,

\be
{\rm Re} f_{\gamma N}(E)=f_{\gamma N}(0)+\frac{E^2}{(2\pi)^2}
P\int_{E_{th}}^{\infty}dE^{\prime}
\frac{\sigma_{\gamma N}(E^{\prime})}{{E^{\prime}}^2 -E^2},
\label{disp}
\ee
where $P$ denotes principle value, $\sigma_{\gamma N}(E)$ is the total
photoproduction cross section, $E_{th}=\mu +\mu^2/2m_N$, $\mu$ and $m_N$
are the pion and nucleon masses, and $f_{\gamma N}(0)$ is given by
the Thompson formula, $f_{\gamma p}(0)=-\alpha/m_p$, $f_{\gamma n}=0$.

The VDM relation Eq.~(\ref{vdm}) holds only for the amplitude of
transversally polarized vector meson $f_{\rho N}^T$,
since $f_{\gamma N}$ is the
scattering amplitude of the real transversally polarized photon. In
Eq.~(\ref{vdm}) the $\rho$-meson energy $E_{\rho}$ is related to the photon
energy by the requirement that the masses of hadronic states produced in
$\rho N$ and $\gamma N$ scattering should be equal,
$E_{\rho}=E_{\gamma}-m_{\rho}^2/2m_N$.

It is known that VDM works well starting from $\gamma$ energies about $2$
GeV, where one may expect the VDM accuracy of about $30\%$ and better at
higher energies (see, e.g.\cite{16}). At these energies the nucleon Fermi
motion can be neglected. In calculation of ${\rm Re} f_{\gamma N}(E)$
according to Eq.~(\ref{disp}) we used the PDG data\cite{17} on photoproduction
on deutron. For the high-energy tail the Donnachie--Landshoff
fitting formula\cite{18} for $\sigma_{\gamma p}$ was used, and it was
assumed that $\sigma_{\gamma D}/\sigma_{\gamma p}=const$ starting
from $E_{\gamma}=20$ GeV. The results for ${\rm Re} f_{\rho N}^T$ and
$\Delta m_{\rho}^T$ at normal nuclear density
$\rho = (4\pi r_0^3/3)^{-1}$, $r_0 =1.25$ fm, are shown in Fig.~1 as
functions of $E_{\rho}$. The mass shift in the energy region, where our
consideration is valid, $2$ GeV$\la E_{\rho}\la 7$ GeV, is positive
($\rho$ mass increases in nuclear matter) and is of order of $50$ MeV.
However, the condition $|{\rm Re} f|< d\sim 2$ fm is not well fulfilled.
Probably the main effect of interference of different nucleons is screening
and the true values of $\Delta m_{\rho}$ are a bit smaller than our results.

Now consider the longitudinal $\rho$-meson.
In this case, unlike the transverse $\rho$, it is
impossible to relate the forward scattering amplitude of $\rho$ to that of
the real photon, but it is still possible to have such a relation for the
virtual photon. We assume that VDM holds for virtual photons if the photon
virtualities are not large, less or of order of $m_{\rho}^2$. For the
transverse scattering amplitude the generalization of Eq.~(\ref{vdm}) to
the virtual photon is

\be
f_{\gamma N}^T (E_{\gamma},q^2)=4\pi\alpha\sum_{V=\rho ,\omega ,\phi}
\frac{m_V^4}{(q^2-m_V^2)^2}\frac{1}{g_V^2}f_{VN}^T (E_V)
\label{t}
\ee
For the longitudinal scattering amplitude the generalization of VDM has the
form

\be
f_{\gamma N}^L (E_{\gamma},q^2)=4\pi\alpha\sum_{V=\rho ,\omega ,\phi}
\frac{|q^2|m_V^2}{(q^2-m_V^2)^2}\frac{1}{g_V^2}f_{VN}^L (E_V)
\label{l}
\ee
Eqs.~(\ref{t}) and (\ref{l}) can be proved in models incorporating direct
$\gamma N$ interaction. The denominators in these equations correspond to
the assumption that at $Q^2=-q^2\la m_V^2$ the dominant intermediate states
in the $\gamma$-channel are vector mesons and the contributions of higher
states can be neglected. The factor $q^2$ in the numerator of Eq.~(\ref{l}) is
a kinematical factor that evidently follows from the requirement of vanishing
$f_{\gamma N}^L$ at $q^2=0$. The absolute value $|q^2|$ arises, since
${\rm Im} f_{\gamma N}^{L}$ is positive at $q^2< 0$ as well as at $q^2> 0$.
This corresponds to the fact that while for transverse photon (or any
transverse or longitudinal vector meson) the polarization vector
squared is $e^2=-1$, for longitudinal virtual photon
we put $e^2=1$ in order to get a
positive cross section (see\cite{16}). The relation between $E_{\rho}$ and
$E_{\gamma}$ is now $E_{\rho}=E_{\gamma}-(m_{\rho}^2 +Q^2)/2m_N$.

Of course, the accuracy in determination of ${\rm Re} f_{\rho N}(E)$
basing on the data for the virtual photon scattering amplitude will
be worse than in the case of real photon, but for the longitudinal
$\rho$-meson even such information will be valuable.
${\rm Re} f_{\gamma N}^{T,L}(E,Q^2)$ can be found from the data on
deep inelastic scattering in the same way as was done for
the real photon. The dispersion relation takes the form

\be
{\rm Re} f_{\gamma N}^{(T,L)}(E,Q^2)=f_{\gamma N}^{T,L}(0,Q^2)-
\frac{\alpha}{m_N}P\int_{0}^{1}dx^{\prime}
\frac{1+4m_N^2 x^2/Q^2}{{x^{\prime}}^2 -x^2}F_2 (x^{\prime},Q^2)
\frac{(1,R)}{1+R}
\label{dis}
\ee
where $x=Q^2/2\nu$, $\nu =m_N E$, $F_2(x,Q^2)$ is the nucleon
structure function, and $R=\sigma_L/\sigma_T$ is the ratio of longitudinal to
transverse photon cross sections.

Consider first the case of transverse photon and check whether starting from
the deep inelastic scattering data we can get the values of
${\rm Re} f_{\rho N}^T (E)$ close to those we have already found from
photoproduction. We choose $Q^2 =0.5$ GeV$^2$ and take
$F_2^p(x,0.5$ GeV$^2)$ from the data compilation done by Ji and
Unrau\cite{19}. The ratio $F_2^n/F_2^p$ was taken from\cite{20} for
$x< 0.2$. For $x> 0.2$, where the data at small $Q^2$ are absent, we
assume $F_2^n/F_2^p=0.75$. The information about $R$ at small $Q^2$ is
scarce. Basing on the data from Refs.\cite{20,21} we assume $R_p =R_n
=0.3$.  We also assume that at $Q^2 =0.5\;$GeV$^2$ the subtraction term in
Eq.~(\ref{dis}) is given by the one-nucleon intermediate state, as it takes
place in the Thompson formula. The one-nucleon intermediate state
contributes also to the integral in Eq.~(\ref{dis}). Its total contribution
to Eq.~(\ref{dis}) is

\be
{\rm Re} f_{\gamma N}^T (\nu,Q^2)_{one-nucl}=-\frac{\alpha}{m_N}\left[
F_E^2 (Q^2)+\frac{1}{4}Q^4 G_M^2(Q^2)\frac{1}{\nu^2 -Q^4/4}\right]\; ,
\label{one}
\ee
where $F_E$ and $G_M$ are the nucleon electric Pauli and
magnetic Sachs formfactors. The results of our calculation show that the
shape of the curve for ${\rm Re} f_{\rho N}^T(E_{\rho})$ obtained from the
data at $Q^2=0.5$ GeV$^2$ is similar to the curve
${\rm Re} f_{\rho N}^T(E_{\rho})$
in Fig.~1, but the absolute values are $30-40\%$ smaller. Since the factor
$(Q^2+m_{\rho}^2)^2/m_{\rho}^4\approx 3.4$ connecting the values of
$f_{\gamma N}^T(E_{\gamma},Q^2)$ and $f_{\rho N}^T(E_{\rho})$ is rather
large, this fact can be considered as an indication that the accuracy of VDM
for the problem considered is of order $30-40\%$.

The calculation of ${\rm Re} f_{\gamma N}^L (E,Q^2)$ is similar. The only
difference appears in the subtraction term in Eq.~(\ref{dis}). In\cite{22}
it was proved that $f_{\gamma N}^L(0,Q^2)$ at small $Q^2$ is given by the
one-nucleon intermediate state and it was argued that its contribution
dominates up to $Q^2=0.5\;$GeV$^2$. The contribution of one-nucleon
intermediate state to $f_{\gamma N}^L (\nu,Q^2)$ is

\be
{\rm Re}f_{\gamma N}^L (\nu,Q^2)_{one-nucl}=-\alpha m_N Q^2\left[
\frac{1}{4m_N^4}F_M^2(Q^2)+\frac{1}{\nu^2 - Q^4/4}G_E^2(Q^2)\right]\; ,
\label{one1}
\ee
where $F_M$ and $G_E$ are the nucleon magnetic Pauli and electric Sachs
formfactors.

The results of calculation of ${\rm Re} f_{\rho N}^L(E_{\rho})$ and
$\Delta m_{\rho}^L(E_{\rho})$ are plotted in Fig.~1.
As is seen from Fig.~1 in
the energy range $E_{\rho}=2-7$ GeV $\Delta m_{\rho}^L$ is essentially
smaller than $\Delta m_{\rho}^T$. Although the uncertainty in the
determination of $\Delta m_{\rho}^L$ is rather large, we believe that this
qualitative conclusion will be intact in a true theory. Since at rest
$\Delta m_{\rho}^T=\Delta m_{\rho}^L$, one should expect a strong energy
dependence of $\Delta m_{\rho}^T$ and/or $\Delta m_{\rho}^L$ in
the domain $m_{\rho}< E_{\rho}< 2$ GeV.  This is not
surprising in the framework of our approach, since there are resonances in
this domain and strong variations of ${\rm Re} f_{\rho N}(E_{\rho})$ and
$\Delta m_{\rho}(E_{\rho})$ are very likely. The main sources of
uncertainty in our approach are the assumption of independent scattering on
nucleons in the nucleus (Fermi gas approximation) and the use of VDM,
especially for the virtual photon. We estimate the uncertainty as $\sim
30-50$\% for $\Delta m_{\rho}^T$ and as a factor of $\sim 2$ for $\Delta
m_{\rho}^L$. The broadening of the $\rho$ width calculated according to
Eq.~(\ref{w}) is large:  $\Delta\Gamma_{\rho}^T\approx 300$ MeV,
$\Delta\Gamma_{\rho}^L\approx 100$ MeV at $E_{\rho}=3$ GeV
and normal nuclear density.

A few remarks are in order comparing our consideration with the previous
ones. Strictly speaking no direct comparison can be done, since all previous
calculations refer to the mass shift of $\rho$-meson at rest, while the
applicability domain of our results is $E_{\rho}> 2$ GeV. As was mentioned
above, one may expect a strong energy dependence of $\Delta m_{\rho}(E)$ in
the interval $m_{\rho}< E_{\rho}< 2$ GeV. (Even the sign difference in
$\Delta m_{\rho}$ obtained here and in Refs.\cite{9,10,13} cannot be
considered as a contradiction, since ${\rm Re} f_{\rho N}(E)$ may change sign
going through resonances, as it indeed
happens with ${\rm Re} f_{\gamma N}(E)$.)
But we would like to emphasize one important point. The basic
physical content of our approach is the statement that the meson mass shift
in nuclear matter is determined by the meson-nucleon interaction and
scattering proceeding at rather large distances, not much less than
internucleon distances. The main point of
Refs.\cite{3,4,5,6,9,10,11,12,13} was the assumption that the mass shifts
are determined by small distances and that the QCD sum rule method developed
for the calculation of small distance contributions can be applied to this
problem. Since our basic formula is general and contains no assumptions
(apart from the Fermi gas approximation, which is a common point in all
approaches) the values of $f_{\rho N}\sim 1$ fm obtained above clearly
demonstrate that large distances are indeed of importance in this problem.
In the calculations of Refs.\cite{6,10,12,13}
the operator product expansion (OPE) for the virtual photon - nucleon
forward scattering amplitude was used
and a few terms in OPE were kept. As is well known the OPE in this case is a
light-cone expansion, and the expansion parameter along the light-cone is
$1/x=2\nu/Q^2$. For the $\rho$-meson at rest $\nu\sim m_N m_{\rho}$,
$Q^2\sim m_{\rho}^2$, and $1/x\sim 2m_N/m_{\rho}\approx 2.5\,$. Therefore,
there are no reasons to keep only a few terms in this expansion, as was done
in\cite{6,10,12,13}. This fact, of course, is the manifestation of the
physical statement made above about importance of large distances in the
problem discussed.

Finally, we would like to mention that a similar treatment of in-medium
pions using the data on $\pi N$ forward scattering amplitudes extracted
from the phase analysis in Ref.\cite{23}
shows a strong energy dependence of the pion mass shift for
$400$ MeV$<E_{\pi}<1500$ MeV: $\Delta m_{\pi}=30-70$ MeV for normal
nuclear density. \vspace{2mm}

This work was supported in part by INTAS Grant 93-0283.

\begin{figure}
\caption{Energy dependence of $-{\rm Re} f_{\rho N}^T$ and
$-{\rm Re} f_{\rho N}^L$ (upper and lower solid curves, left scale), and
of $\Delta m_{\rho}^T$ and $\Delta m_{\rho}^L$ (upper and lower dashed
curves, right scale) at normal nuclear density.}
\end{figure}
\end{document}